\documentclass[epsf,prb,twocolumn,showpacs,preprintnumbers]{revtex4}
\usepackage{graphics}
\usepackage{graphicx}
\usepackage{dcolumn} 
\usepackage{bm}
\usepackage{color}
\usepackage{epsfig}
\newcommand{\boo}{BaOsO$_3$}
\newcommand{\bro}{BaRuO$_3$}
\newcommand{\coo}{CaOsO$_3$}

\begin{document}
\title{Electronic Structures, Magnetism, and Phonon Spectra 
 in the Metallic Cubic Perovskite BaOsO$_3$
}
\author{Myung-Chul Jung$^1$} 
\author{Kwan-Woo Lee$^{1,2}$}
\email{mckwan@korea.ac.kr}
\affiliation{ 
 $^1$Department of Applied Physics, Graduate School, Korea University, Sejong 339-700, Korea\\
 $^2$ Department of Display and Semiconductor Physics, 
  Korea University, Sejong 339-700, Korea
}
\date{\today}
\pacs{71.20.Be, 71.20.Dg, 75.50.Cc, 71.18.+y}
\begin{abstract}
By using {\it ab initio} calculations, we investigated a cubic perovskite BaOsO$_3$ 
and a few related compounds that have been synthesized recently and 
formally have a metallic $d^4$ configuration. 
In BaOsO$_3$, which shows obvious three-dimensional fermiology, 
a nonmagnetism is induced by a large spin-orbit coupling (SOC),
which is precisely equal to an exchange splitting $\sim$0.4 eV of the $t_{2g}$ manifold.
However, the inclusion of on-site Coulomb repulsion as small as $U^c\approx$1.2 eV, 
only $1/3$ of the $t_{2g}$ bandwidth, leads to the emergence of a spin-ordered moment, 
indicating that this system is on the verge of magnetism.
In contrast to BaOsO$_3$, our calculations suggest that
the ground state of an orthorhombic CaOsO$_3$ is a magnetically ordered state 
due to the reduction of the strength of SOC (about a half of that of BaOsO$_3$)  
driven by the structure distortion, although the magnetization energy is only 
a few tenths of meV.
Furthermore, in the cubic BaOsO$_3$ and BaRuO$_3$, our full-phonon calculations
show several unstable modes, requiring further research.

\end{abstract}
\maketitle

\section{Introduction}
Transition metal oxides show abundant phenomena involved in the interplay
among charge, spin, orbital, and lattice degrees of freedom.
Recently, the effects of spin-orbit coupling (SOC) have been intensively investigated in
the $t_{2g}$ manifold of an {$\cal M$}O$_6$ octahedral structure 
({$\cal M$=$5d$ transition metals}).\cite{d1_LP,d1_kim,d2_chen,d3_LP1,d3_LP2,d3_mat,d5_yu,d5_kim,d5_liu,d4_kha,d4_meet}
In the atomic limit, a large SOC leads to transforming the $t_{2g}$ manifold
into an effective angular momentum of $|{\cal L}|$=1.\cite{d1_LP} 


A conventional SOC picture says that a $d^4$ system is a trivial nonmagnetic insulator
in the large SOC limit, which may be suitable for a nonmagnetic (NM) 
and nonmetallic NaIrO$_3$.\cite{cava}
However, two interesting studies have recently appeared.
Khaliullin suggested a van Vleck-type Mott insulating state due to excitations
between a singlet $J=0$ state and triplet $J=1$ states.\cite{d4_kha}
Meetei {\it et al.} proposed unusual Mott insulators of 
a charge-disproportionated ferromagnetic (FM) $J=1$ state as well as a FM $J=2$ state,  
as including effects of correlation.\cite{d4_meet}
One may expect different physical phenomena for a metallic $5d^4$ system such as BaOsO$_3$,
which will be focused on in this research.

Several decades ago, Chamberland and Sarkozy synthesized a body-centered cubic KSbO$_3$-type
and a $6H$ hexagonal phase.\cite{cham73,cham78}
Although no detailed information of the crystal structures and the physical properties is available, 
the $6H$ hexagonal phase, which is a semiconductor with an activation energy of 0.39 eV,
follows the Curie-Weiss behavior above 100 K (no measurement has been done below this temperature). 
The observed effective moment is 2.81 $\mu_B$, which is very close to the spin-only value for $S=1$, 
indicating negligible effects of SOC for such a large structure distortion.\cite{cham78}
Very recently, Shi {\it et al.} synthesized a distortion-free cubic perovskite phase 
using a technique of high temperature ($\sim$2000 K) and high pressure (17 GPa).\cite{boo}
The resistivity is metallic but indicates a small upturn at 50 K, which may be 
due to the polycrystallinity of the sample.
The heat capacity measurement shows a metallic behavior, but 
can be fitted well by a linear combination of the Debye and the Einstein models, 
suggesting a complicated behavior in phonon modes around 300 K.\cite{boo}
Shi {\it et al.} also synthesized two orthorhombic, isovalent perovskites \coo~
and SrOsO$_3$.\cite{boo}
\coo~ shows complicated electronic properties, whereas SrOsO$_3$ has metallic characteristics
in both the resistivity and the heat capacity measurements.
The resistivity data of \coo~ is semiconductor-like, but $\ln \rho(T)$ does not follow
typical models of $T^{-1}$ or $T^{-1/4}$ forms.
Also, the heat capacity contains itinerant electronic character
even at very low $T$. It was claimed that \coo~ is near a Mott insulating state.\cite{boo}
For the magnetic properties,
these three osmates commonly follow the Curie-Weiss susceptibility, though
no order moment has been observed.
At low T, the susceptibility measurements of \coo~ and SrOsO$_3$ show  
larger enhancement than that of \boo, 
indicating more magnetic tendencies in these two orthorhombic osmates. 

Through first-principles calculations, we will address 
the electronic structures and magnetic properties of the cubic \boo, 
which are substantially affected by SOC, structure distortion, and correlation effects.
The cubic \boo~ will be compared and contrasted to the orthorhombic \coo~ 
to investigate the interplay between magnetism, SOC, and structure distortion.
Besides, we performed full-phonon calculations to inspect the stability
of the cubic phases of isovalent \boo~ and BaRuO$_3$,
which can be synthesized only 
by a technique of very high pressure and temperature.\cite{boo,bro}

\section{Calculation Method}
Our calculations were carried out by using the experiment lattice parameter $a=4.02573$ \AA~ 
for \boo \cite{boo}~ with two all-electron full potential codes, {\sc fplo}\cite{fplo1} 
and {\sc wien2k}\cite{wien2k}. 
Both SOC and correlations have been considered within extensions of 
the Perdew-Burke-Erznerhof generalized gradient approximation (PBE-GGA) functional.\cite{pbe}
When necessary, we optimized the lattice or internal parameters until the forces were smaller than
1 meV/a.u. (see below).
The Brillouin zone was sampled with a very dense $k$-mesh
of 24$\times$24$\times$24 to check the convergence carefully for this metallic system. 
In  {\sc wien2k}, the basis size was determined by $R_{mt}K_{max}=7$ 
and the augmented plane wave (APW) radii of Ba 2.50, Os 2.08, O 1.7 (in units of a.u.). 

We also performed linear response phonon calculations, using {\sc quantum espresso}\cite{qe} 
with PBE-GGA ultrasoft-pseudopotential\cite{pbe} and the fully relativistic pseudopotential for GGA+SOC. 
These calculations were carried out with a 4$\times$4$\times$4 $q$-mesh, 
a 24$\times$24$\times$24 $k$-mesh, an energy cutoff of 45 Ry, 
and a charge cutoff of 450 Ry.

\begin{figure}[tbp]
{\resizebox{8cm}{6cm}{\includegraphics{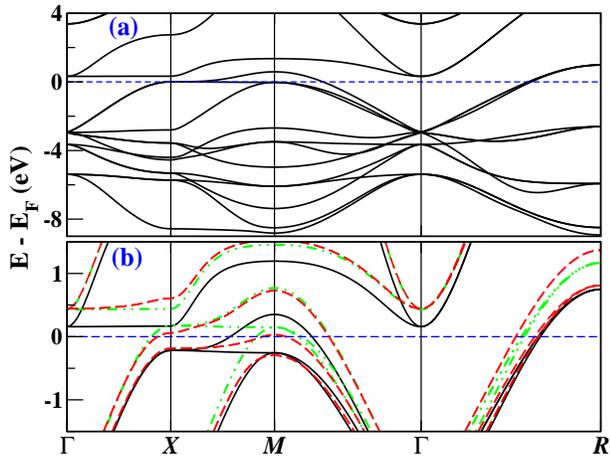}}}
\caption{(Color online) (a) GGA NM band structure of \boo~ in the regime
including O $2p$ and Os $5d$ orbitals.
The bands above 0.5 eV are the Os $e_g$ manifold.
(b) Blowup GGA FM band structure (black solid lines are for spin up and 
green dot-dashed lines are for spin down), 
which is overlapped by the GGA+SOC one (red dashed lines).
The horizontal dashed lines denote the Fermi energy $E_F$, which is set to zero.
}
\label{band}
\end{figure}

\begin{figure}[tbp]
{\resizebox{8cm}{7.5cm}{\includegraphics{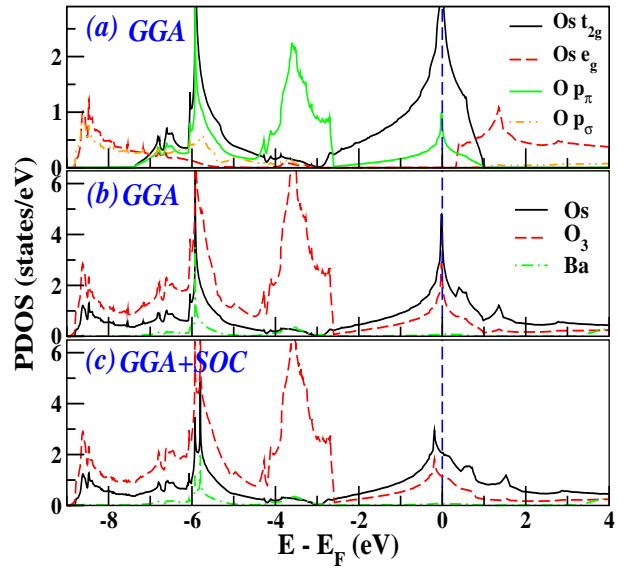}}}
\caption{(Color online) Orbital-projected densities of states (PDOSs) per atom in (a) GGA, 
and (b) atom PDOSs in GGA and in (c) GGA+SOC of NM \boo.
The corresponding FM DOS is not shown here, since this is very similar to that of NM,
except for the positions of $E_F$'s at 0.25 eV and --0.15 eV 
for the spin-up and -down channels, respectively, relative to $E_F$ of NM.
For NM, the single spin DOS at $E_F$ is $N(E_F)=$3.38 states per eV.
}
\label{pmdos}
\end{figure}

\section{Results}
\subsection{Electronic Structure of BaOsO$_3$}
First, we address the electronic structure of nonmagnetic (NM) \boo, 
since the ground state is NM when considering SOC. 
We discuss this in Sec. \ref{magnet} in detail.

The top of Fig. \ref{band} displays the NM band structure of \boo, excluding SOC.
The O $p$ states and the Os $t_{2g}$ manifold extend over the range of --9 eV to 0.5 eV,
relative to the Fermi energy $E_F$.
As clearly visible in the corresponding DOS of Fig. \ref{pmdos}(b),
$E_F$ pinpoints at the $2/3$ filling of the $t_{2g}$ manifold, 
which is consistent with the $d^4$ configuration. 
The orbital-projected DOS of Fig. \ref{pmdos}(a) shows
strong hybridization between Os $5d$ and O $2p$ states: roughly
$pd\sigma$ states at --9 eV to --5 eV (0.5 eV to 6 eV), $pd\pi$ states
 at --7.2 eV to --3 eV (--3 eV to 1 eV), and pure O $p_\pi$ states at --4.2 eV to --2.8 eV.
Here, the numbers inside the parentheses are for each antibonding states, 
usually denoted the $e_g$ and $t_{2g}$ manifolds, respectively.
These bands below 1 eV were fitted by the tight-binding Wannier function technique
implemented in {\sc fplo}.
We obtained two important parameters: $pd\pi$ hopping $t_\pi$=1.32 eV and 
direct oxygen-oxygen hopping $t_\pi'$=0.14 eV.\cite{mazin}
The $t_\pi'$ value is similar to that of \bro, but the value of $t_\pi$ is about 10\% larger than
that of \bro,\cite{KU_bro} reflecting the wider extension of the $5d$ orbital than 
the $4d$ orbital. 
Remarkably, in the band structure, there are three nearly dispersionless bands 
around the $M$ point, in addition to a flat band just above $E_F$ 
along the $\Gamma-X$ line, which often appears 
due to a lack of $dd\delta$ hopping in conventional perovskites.\cite{mgcni3}
The former three bands lead to sharp peaks at --6 eV, --3.5 eV, and $E_F$ 
in the DOS of Fig. \ref{pmdos}(b), 
whereas the latter conduction band does not.
The van Hove singularity at $E_F$ results from the quite flat band
with the character of $pd\pi$ antibonding between Os $d_{xz}$ and O $p_z$ orbitals
along the $X-M$ line, which shows a large band mass of $m^\ast \approx-70$.  
This van Hove singularity enhances the magnetic instability (see below).

To investigate the effects of SOC, we carried out GGA+SOC calculations.
The blowup band structure near $E_F$ denoted 
by (red) dashed lines in the bottom of Fig. \ref{band} shows
a strength of SOC of $\xi \approx0.4$ eV, as measured near $E_F$, 
where SOC affects considerably.
This value is largely reduced from the atomic value $\sim$1 eV 
due to the strong $p-d$ hybridization.
Including SOC, the flat $pd\pi$ antibonding band becomes more dispersive
and shifts down, leading to making the van Hove singularity blunted below $E_F$,
as shown in the PDOS in Fig. \ref{pmdos}(c). 
This system becomes very three dimensional.
As a result, the total DOS $N(E_F)$ becomes only a half of that of the GGA NM,
suggesting that the magnetic instability is significantly diminished by SOC, 
as discussed below.

\begin{figure}[tbp]
{\resizebox{8cm}{12cm}{\includegraphics{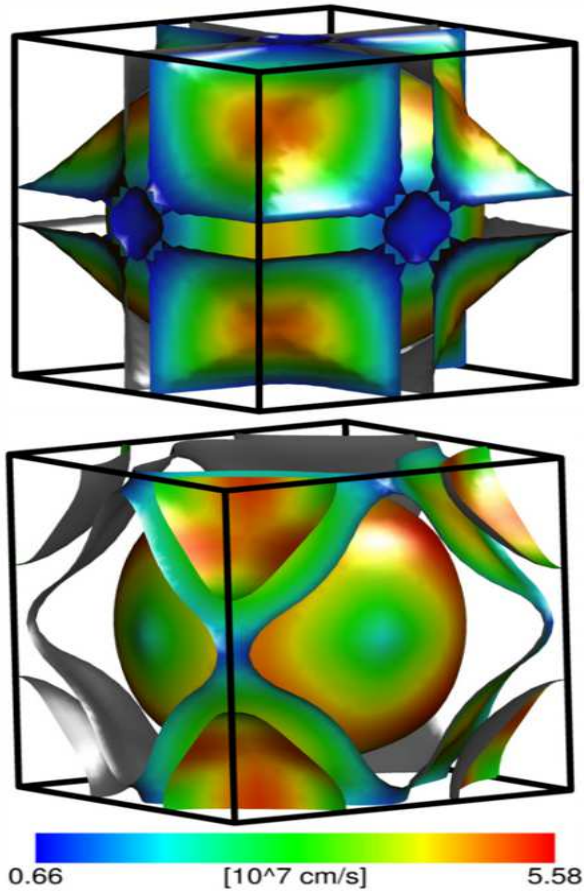}}}
\caption{(Color online) NM Fermi surfaces (FSs) in both GGA (top panel) and GGA+SOC (bottom panel) 
colored with Fermi velocity, which show three-dimensional character.
In the GGA case the $R$-centered hole FS is not shown for better visualization, 
since this FS is very similar to that of GGA+SOC, except for a little larger size.
The $R$ and $M$ points are at each corner and midway of each side, respectively.
}
\label{fs}
\end{figure}

\subsection{Fermiology}
The NM Fermi surfaces (FSs) colored with the Fermi velocity are displayed
in Fig. \ref{fs}.
FSs of both GGA and GGA+SOC consist of three pieces:
two hole-like pieces and a $\Gamma$-centered spherical electron-like piece.
Neglecting SOC, FSs have more flat regimes in various directions,
implying the possibility of a charge or spin fluctuation.
The FSs have much lower Fermi velocity in the $\langle 100\rangle$ directions,
reflecting the fact that bands at $E_F$ are quite dispersionless in the direction.
However, our trials to obtain antiferromagnetic states
of three typical types are always converged to NM or FM states 
in both GGA and GGA+SOC.
This would rule out the possibility of a spin fluctuation.

Considering the effects of SOC, the most significant change occurs 
in the open FS surrounding the $\Gamma$-centered sphere, 
resulting in being an $M$-centered sandglass-like shape.
In the FS, most flat parts disappear, but some survive in faces toward the $R$ point.
The $R$-centered FS is shrunk and becomes nearly spherical
with a radius of $\sim2/3(\pi/a)$, containing $\sim$0.31 holes.
The $\Gamma$-centered spherical FS becomes more isotropic, although
nodules appear in the $\langle 100\rangle$ directions. 
The radius of the spherical FS decreases by $\sim$10\% 
to compensate for the reduction of the hole-like FSs,
and attains a radius of 0.9$\pi/a$, containing about 0.76 electrons.

Transport properties can be studied with the Fermi velocity and plasma frequency. 
For the GGA case, the root-mean-square of the Fermi velocity is
$v_F^{rms}=2.71\times10^7$ cm/s, a typical value for a metal.
The plasma energy $\hbar\Omega_{p,ii}$, which is proportional to $v_F^{rms}\times \sqrt{N(E_F)}$,
is 4.55 eV. 
The inclusion of SOC increases $v_F^{rms}$ to $3.63\times10^7$ cm/s, 
which is consistent with the fact that the bands around $E_F$ become more dispersive. 
So, the magnitude of the plasma energies is similar to the case neglecting SOC.

\begin{figure}[tbp]
{\resizebox{8cm}{6cm}{\includegraphics{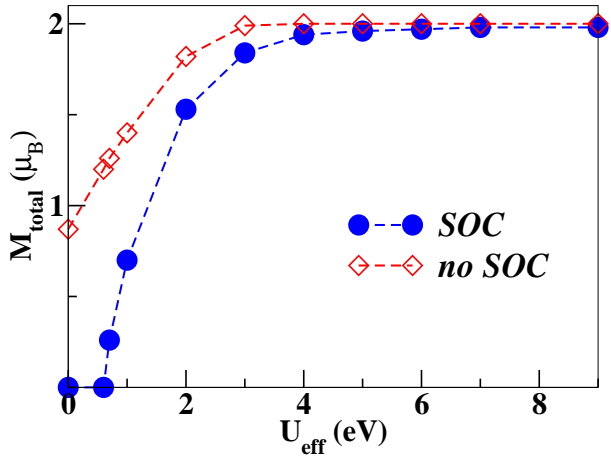}}}
\caption{(Color online) Total spin moment versus strength of the effective on-site Coulomb
 repulsion $U_{eff}$ in GGA+U+SOC.
 For comparison, results of GGA+U are also given.
 In GGA+U+SOC, a small moment appears at $U_{eff}^c=0.6$ eV and
 is saturated to $\sim$2$\mu_{B}$ above $U_{eff}\approx4$ eV.
 For almost all range of $U_{eff}$ studied here, the Os orbital moment is
 negligible, but 0.1 $\mu_B$ to 0.2 $\mu_B$ appears above $U_{eff}=6$ eV due to
 an enhanced orbital-polarization in the large $U$ regime.
}
\label{mu}
\end{figure}

\subsection{Magnetic tendencies}
\label{magnet}
Within GGA, FM with a total moment of 0.85 $\mu_B$  
is energetically favored over NM, but the difference in energy is only 11 meV.
We performed fixed spin moment calculations in GGA to investigate
the stability of the FM state.
We obtained $I=$0.79 eV, leading to $IN(E_F)\approx1.2$ with $N(E_F)=1.53$ states/eV-spin 
for FM. This value, above unity of the Stoner criterion, 
is close to that of the cubic FM \bro~ with the similar moment.\cite{KU_bro} 

The bottom of Fig. \ref{band} displays the enlarged FM band structure in GGA,
showing that the exchange splitting of the $t_{2g}$ manifold
is about 0.4 eV.
Remarkably, this value is identical to the strength of SOC,
resulting in a transition from FM to NM in this metallic $d^4$ system.
Whenever including SOC,\cite{soc} our trials always converge to NM,
which is consistent with the experiment.

Although the strength of correlation is still unclear in this system, 
the linear specific coefficient ratio of $\gamma_{exp}/\gamma_0$ 
can be a guide to determine whether correlation is considered.
From $N(E_F)$ obtained from GGA+SOC, $\gamma_0=$7.68 (in units of mJ/mol-K$^2$),
which is consistent with a previous report.\cite{boo}
Compared with the experiment value of $\gamma_{exp}=16.8$,\cite{boo} 
the ratio of $\gamma_{exp}/\gamma_{0}\approx$ 2 implies a moderate correlation strength
in \boo.
So, using the GGA+U+SOC approach, we applied the effective on-site Coulomb repulsion 
$U_{eff} = U - J$
to the Os ion to inspect whether this system is close to a magnetic ordering. 
Here, $J$ is Hund's exchange integral.
The change in the total spin moment is described with varying strength of $U_{eff}$ in Fig. \ref{mu}.
At $U_{eff}$ as small as 0.6 eV, a magnetic moment emerges.
Assuming a typical $J\approx0.6$ eV for $5d$ systems, the critical on-site Coulomb repulsion 
$U^c$ is only 1.2 eV, which is obviously within the range of a reasonable $U$ value 
for heavy transition metals.
This fact suggests that this system is in the vicinity of a spin-ordered state,
which is consistent with the Curie-Weiss behavior in the susceptibility.\cite{boo}
Also, as $U_{eff}$ increases, the moment rapidly increases and reaches 
to the saturated value of $\sim$2 $\mu_B$ at $U_{eff}\approx4$ eV.
However, this system remains metallic even for a very large $U_{eff}=9$ eV 
in GGA+U+SOC, whereas a half-metallic state appears above $U_{eff}=4$ eV 
for the case neglecting SOC.

\begin{figure}[tbp]
{\resizebox{8cm}{6cm}{\includegraphics{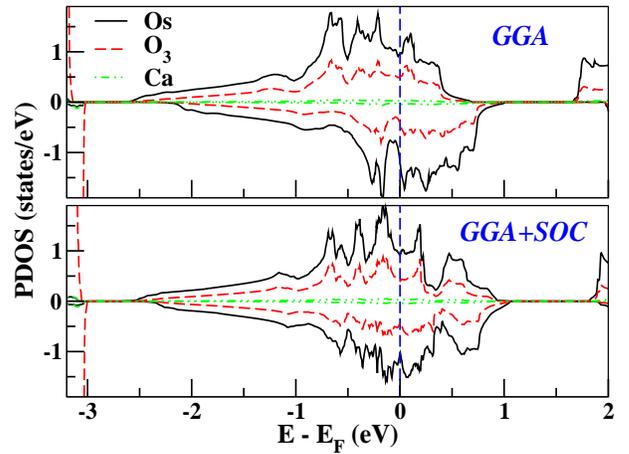}}}
\caption{(Color online) Enlarged PDOSs in GGA (top panel) and in GGA+SOC (bottom panel) 
of FM CaOsO$_3$ in the region of the partially filled $t_{2g}$ manifold,
which is 25\% narrower than in \boo.
$N(E_F)$ is 4.40 (3.71) and 3.16 (2.82) in NM and FM, respectively, in units of states/eV
per formula unit.
The values inside parentheses are obtained from GGA+SOC calculations.
}
\label{coo_dos}
\end{figure}

\subsection{Interplay among structure distortion, SOC, and magnetism}
 A structure distortion of the ${\cal M}$O$_6$ octahedron from the ideal one
reduces the strength of SOC.  
So, one may expect a magnetic ordering to emerge in the orthorhombic perovskite CaOsO$_3$
with relatively small structure distortion,\cite{boo}
since the NM ground state of the cubic \boo~ is purely due to effects of SOC.
In GGA, FM with the moment of 0.82 $\mu_B$/Os has lower energy by 6 meV/f.u. than NM,
which is consistent with our fixed spin moment calculations.
The obtained Stoner parameter $I=0.76$ eV is a little smaller than in \boo,
but $IN(E_F)$ is almost identical to that of \boo.
This indicates similar magnetic tendencies in both CaOsO$_3$ and \boo, when ignoring SOC.

Figure \ref{coo_dos} shows the blowup PDOS in both GGA and GGA+SOC,
which indicate the exchange splitting of the $t_{2g}$ manifold of
about 0.3 and 0.1 eV for GGA and GGA+SOC, respectively.
This reduction of 0.2 eV is almost identical to the strength of SOC in CaOsO$_3$,
which is only a half of that of \boo.  
Consistent with this reduction, the inclusion of SOC reduces the moment by 50\% and 
leads to a little magnetization energy, by only a few tenths of meV,
although FM remains favored.
A Stoner parameter in GGA+SOC can be estimated by $I\sim 1/N(E_F)=0.35$,
which is a half of the GGA value.
Our results suggest that the unusual magnetism, observed in the experiment, in the orthorhombic CaOsO$_3$
results from emerging magnetic ordering induced by reducing the strength of SOC.

\begin{figure}[tbp]
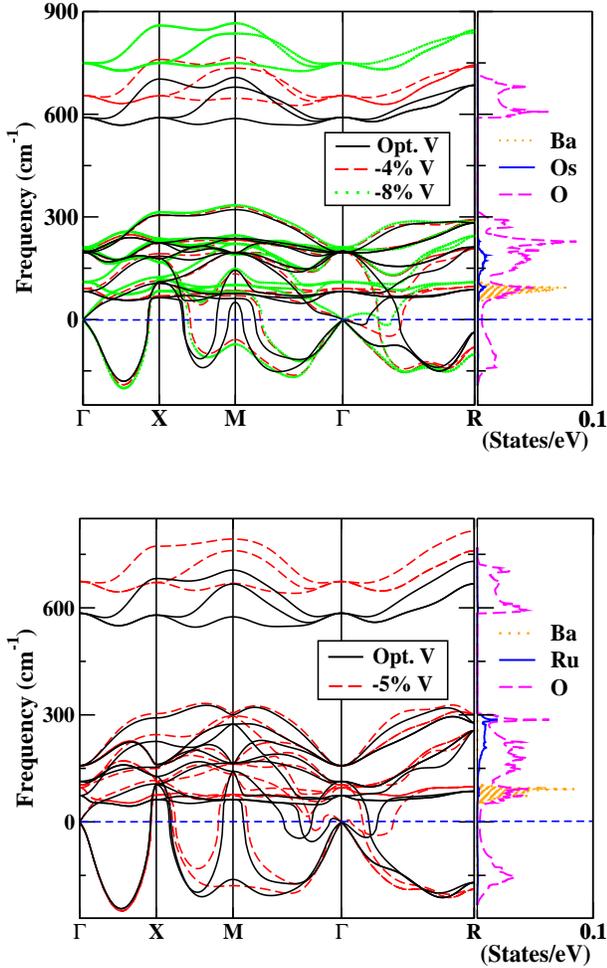

{\resizebox{8cm}{6cm}{\includegraphics{Fig6a.eps}}}
\vskip 8mm
{\resizebox{8cm}{6cm}{\includegraphics{Fig6b.eps}}}
\caption{(Color online) Top: Cubic \boo~ phonon dispersion curve and atom-projected phonon DOSs,
obtained from GGA+SOC by using our optimized volume 
and the 4\% and 8\% reduced volumes from the optimized value.
Bottom: The same plots of cubic BaRuO$_3$ in GGA with our optimized 
and 5\% reduced volumes. The latter corresponds to the experimental value.
SOC is excluded, since the strength of SOC is small.\cite{KU_bro}
In both plots, each atom-projected phonon DOS is given for the optimized volume.
Negative values in the phonon dispersion curves correspond to imaginary phonon frequencies.  
}
\label{phonon}
\end{figure}
 
\subsection{Phonon instability}
Through first principles calculations, Rahman {\it et al.} suggested 
that two types of Jahn-Teller distortions in a cubic and isovalent BaFeO$_3$, 
which are related to an antiferroic distortion of either Os or O ions along the $c$-axis,
lead to a magnetic phase transition.\cite{bfo_ph}
In the cubic \boo, our calculations show that these distortions are energetically unfavored,
which is consistent with the phonon dispersion (see below). 
As mentioned in the introduction, however, the measurement of the specific heat suggested
an unusual behavior.\cite{boo}
Thus, we carried out linear response phonon calculations for our GGA optimized
and a few compressed volumes, using both GGA and GGA+SOC.

Within GGA, we obtained several imaginary frequencies throughout all regimes (not shown here).
Even for the inclusion of SOC, most of these unstable modes still survive, 
as given in the top of Fig. \ref{phonon}. 
As often observed in cubic perovskites,\cite{srzro3} there are 
unstable modes, corresponding to the tilting of oxygens (triplet $R_{25}$, 36$i$ cm$^{-1}$)
and the rotation of in-plane oxygens (singlet $M_3$).\cite{cowley}
The latter appears at $\sim60i$ cm$^{-1}$ in the compressed volume. 
Structure distortions involving these modes have been intensively investigated
in several perovskites by Amisi {\it et al.} and He {\it et al.},\cite{srzro3,srpdo3} 
who suggest the gadolinium orthoferrite GdFeO$_3$-type structural distortion 
as the stablest one.
However, in \boo~ the GdFeO$_3$-type distortion is energetically unfavored.
The most unstable modes of doublet 180$i$ cm$^{-1}$ appear midway of the $\Gamma$--$X$ line, 
and are involved in the transverse acoustic mode, in which 
a displacement of one of the planar oxygens along the [110] direction is dominant.
There are a few additional unstable modes along the $M$--$\Gamma$ 
and the $\Gamma$--$R$ lines. The latter is similar to 
the instability of the transverse acoustic mode observed in 
the possible charge-density-wave (CDW) Y$_3$Co.[\onlinecite{y3co}]
However, consistent with the strong three-dimensionality in \boo,
no indication of CDW is observed in our calculations.
We also performed these calculations with two compressed volumes, 
which are 4\% and 8\% smaller than the optimized value, 
to investigate effects of pressure on the instability.
The former corresponds to the experimental value.
As shown in the top of Fig. \ref{phonon}, most of the unstable modes 
remain nearly unchanged,
indicating that volume contraction does not play an important role in the instability.
The complicated phonon instability seems to be of interest, requiring
further studies in both theoretical and experimental viewpoints, 
although this is not an issue covered in this research.

For comparison, phonon calculations were carried out in the cubic BaRuO$_3$,
which is cubic even at temperature as low as 10 K.\cite{bro}
The bottom of Fig. \ref{phonon} displays our results, which show very similar
behavior to that of \boo.
It is unclear yet why these phonon spectra show serious instabilities, 
although both the cubic \boo~and BaRuO$_3$ phases are experimentally stable.
One possible scenario is that quantum fluctuations would stabilize 
the cubic phase in both systems, as discussed for some cubic perovskites.\cite{vand,singh}
This is true, only when the difference in energy between the cubic phase and a distorted phase 
is very small.   
Also, the instability may be related to the unusual behavior in the specific heat and
the fact that these cubic systems can be synthesized only by a technique of extreme conditions.

\section{Discussion and Conclusion}
Comparing the experimental linear specific heat coefficient $\gamma_{exp}$=18 mJ/mol-K$^2$ of \coo~ 
with our theoretical values, there is an enhancement of a factor of two to three,
implying a moderate correlation.
Within the GGA+SOC+U approach, applying $U_{eff}$ to Os ions in \coo~ leads 
to an insulating state at $U_{eff}^c\approx$6.8 eV via a half-metallic state
at $U_{eff}\approx$4 eV.
It may imply that this system is near an insulating state,
resulting in the atypical electrical properties of \coo.

In conclusion, we have investigated the effects of SOC on the magnetism 
of a few metallic $5d^4$ systems synthesized recently.
The cubic \boo~ shows a transition from a Stoner-type FM to NM
incurred by SOC, of which the strength is identical to the exchange splitting of 
the $t_{2g}$ manifold. However, at small $U^c\approx$1.2 eV 
in GGA+SOC+U calculations, a small magnetic moment revives,
suggesting that \boo~ is an incipient magnet.
On the other hand, our results indicate that the orthorhombic and isovalent \coo~ is FM,
since the structure distortion leads to reducing the strength of SOC.
We anticipate this fact to be true for the isostructural and isovalent SrOsO$_3$. 
These results are consistent with a spin-ordered state observed very recently 
in a distorted double perovskite Sr$_2$YIrO$_6$ with $5d^4$ Ir$^{5+}$ ions.\cite{cao}
Our findings indicate that these systems are good examples to investigate
the interplay between magnetism, lattice, SOC, and correlation.
Further research is required to clarify these issues from both the theoretical
and experimental viewpoints.

\section{Acknowledgments}
We acknowledge P. Giannozzi and C.-J. Kang for useful discussions on phonon instabilities, 
and W. E. Pickett for fruitful communications in the early stage of this research.
This research was supported by NRF-2013R1A1A2A10008946.


\begin{thebibliography}{10}
\bibitem{soc} Our LSDA+SOC calculations show the same results, indicating
that these results are insensitive to a selection of the exchange-correlation
functional.
\end{thebibliography}
\end{document}